\documentclass[sigconf]{acmart}

\AtBeginDocument{%
  \providecommand\BibTeX{{%
    \normalfont B\kern-0.5em{\scshape i\kern-0.25em b}\kern-0.8em\TeX}}}
    
\usepackage{amsmath,amssymb,amsthm,bm,booktabs,comment,multirow}
\newcommand{\Eqref}[1]{Eq.~(\ref{#1})}
\newcommand{\Thmref}[1]{Theorem~\ref{#1}}
\newcommand{\Lemref}[1]{Lemma~\ref{#1}}

\newcommand{\Propref}[1]{Proposition~\ref{#1}}
\newcommand{\Figref}[1]{Figure~\ref{#1}}
\newcommand{\Tabref}[1]{Table~\ref{#1}}
\renewcommand{\refname}{Reference}

\newcommand{\argmin}{\mathop{\rm arg~min}\limits}
\usepackage{algorithm}
\usepackage{algpseudocode}

\copyrightyear{2020} 
\acmYear{2020} 
\setcopyright{acmcopyright}\acmConference[SIGIR '20]{Proceedings of the 43rd International ACM SIGIR Conference on Research and Development in Information Retrieval}{July 25--30, 2020}{Virtual Event, China}
\acmBooktitle{Proceedings of the 43rd International ACM SIGIR Conference on Research and Development in Information Retrieval (SIGIR '20), July 25--30, 2020, Virtual Event, China}
\acmPrice{15.00}
\acmDOI{10.1145/3397271.3401114}
\acmISBN{978-1-4503-8016-4/20/07}

\begin{document}
\fancyhead{}
\title{Asymmetric Tri-training for Debiasing \\
Missing-Not-At-Random Explicit Feedback}

\author{Yuta Saito}
\affiliation{Tokyo Institute of Technology}
\email{saito.y.bj@m.titech.ac.jp}

\renewcommand{\shortauthors}{Y.Saito.}

\begin{abstract}
   In most real-world recommender systems, the observed rating data are subject to selection bias, and the data are thus \textit{missing-not-at-random}. Developing a method to facilitate the learning of a recommender with biased feedback is one of the most challenging problems, as it is widely known that naive approaches under selection bias often lead to suboptimal results. A well-established solution for the problem is using propensity scoring techniques. The propensity score is the probability of each data being observed, and unbiased performance estimation is possible by weighting each data by the inverse of its propensity. However, the performance of the propensity-based unbiased estimation approach is often affected by choice of the propensity estimation model or the high variance problem. To overcome these limitations, we propose a model-agnostic meta-learning method inspired by the \textit{asymmetric tri-training} framework for unsupervised domain adaptation. The proposed method utilizes two predictors to generate data with reliable pseudo-ratings and another predictor to make the final predictions. In a theoretical analysis, a propensity-independent upper bound of the true performance metric is derived, and it is demonstrated that the proposed method can minimize this bound. We conduct comprehensive experiments using public real-world datasets. The results suggest that the previous propensity-based methods are largely affected by the choice of propensity models and the variance problem caused by the inverse propensity weighting. Moreover, we show that the proposed meta-learning method is robust to these issues and can facilitate in developing effective recommendations from biased explicit feedback.
\end{abstract}



\maketitle

\section{Introduction}
The goal of recommender systems is to recommend items that users will prefer. To achieve this, recommendation algorithms predict the potential preference or relevance of non-interacted user-item pairs by using sparse observed ratings. Developing effective recommendation algorithms is critical to improving the profit margin of marketing platforms (e.g., Amazon and Etsy) or the user experience in interactive systems (e.g., Spotify and Netflix). Therefore, the field of personalized recommendation has been widely studied in both the academia and industry.

Within the area of recommender systems, most existing studies assume that the observed rating data are \textit{missing-completely-at-random} (MCAR). Generally, this assumption does not hold because real-world recommender systems are subject to \textit{selection bias}. Selection bias occurs primarily due to the following two reasons~\cite{schnabel2016recommendations,wang2019doubly}. First, the probability of observing each rating is highly dependent on a past recommendation policy. For example, if the observed rating dataset is collected under the most popular policy, a policy that always recommends some of the popular items to all users, the probability of observing ratings of such popular items may be large. This leads to the non-uniform missing mechanism, and the MCAR assumption is violated. Second, user self-selection happens, as users are free to choose the items that they wish to rate. For example, in a movie recommender system, users usually watch and rate movies that they like and rarely rate movies that they do not like~\cite{pradel2012ranking}. Another example is a song recommender system, in which users tend to rate songs that they like or dislike and seldom rate songs they feel neutral about~\cite{marlin2009collaborative}. These findings suggest that most rating datasets collected through real-world recommender systems are \textit{missing-not-at-random} (MNAR). Several studies have theoretically and empirically indicated that the conventional methods of naively using observed ratings lead to suboptimal prediction models, as the observed ratings are not representative data of the target population~\cite{steck2010training,schnabel2016recommendations,yang2018unbiased}. Thus, developing a recommendation algorithm and a debiasing method that can achieve a high prediction accuracy using MNAR feedback is essential to achieve the goal of recommender systems in the real-world.

Several related approaches directly address the MNAR problem. Among these, the most promising approaches are propensity-based debiasing methods such as \textit{inverse propensity score} (IPS) and \textit{doubly robust} (DR) estimations. These methods have been established in fields such as causal inference and missing data analysis~\cite{rosenbaum1983central, rubin1974estimating} and have been proposed to be utilized for debiasing learning and evaluation of the MNAR recommendation~\cite{schnabel2016recommendations, liang2016causal}. IPS estimation relies on the propensity score, which is the probability of observing each rating. By weighting each sample by the inverse of its propensity score, one can unbiasedly estimate the loss function of interest using the biased rating feedback. The benefits of these propensity-based methods are theoretically principled and empirically outperform naive methods based on the unrealistic MCAR assumption~\cite{schnabel2016recommendations,liang2016causal}.

To ensure effectiveness of such propensity-based methods, accurately estimating the propensity score is critical. This is because the unbiasedness of the performance estimator is guaranteed only when the true propensities are available; the IPS estimator still has a bias depending on the propensity estimation bias~\cite{schnabel2016recommendations, wang2019doubly}. However, correctly estimating the propensity score is almost impossible, and model misspecification often occurs in real-world settings~\cite{saito2019doubly}. Moreover, propensity-based methods generally suffer from high variance, which can lead to suboptimal estimation when the item popularity or user activeness is highly diverse~\cite{wang2019doubly, gilotte2018offline, swaminathan2015self}. Improving the robustness to the choice of propensity estimator and the high variance problem of the propensity weighting technique are the important and unsolved issues.

To address the limitations of the propensity-based recommendation methods, in this work, we propose a model-agnostic meta-learning method inspired by the \textit{asymmetric tri-training} framework in unsupervised domain adaptation~\cite{saito2017asymmetric}. Similar to causal inference, unsupervised domain adaptation addresses problem settings in which the data-generating distributions are different between the training and test sets~\cite{kuroki2018unsupervised,lee2019domain}. Moreover, it relies on the upper bound minimization approach, which minimizes the propensity-independent upper bound of the loss function of interest. Thus, this approach is considered to be useful to overcome the issues related to the propensity weighting technique.

In the theoretical analysis, we establish the new upper bound of the ideal loss function and demonstrate that the proposed meta-learning method attempts to minimize this bound. In contrast to the bounds presented in previous studies~\cite{schnabel2016recommendations,wang2019doubly}, the upper bound minimized by the proposed method is independent of the propensity score; thus, issues related to the inverse propensity weighting are expected to be solved. Finally, we conduct extensive experiments using public real-world datasets. In particular, we demonstrate that the performance of the previous propensity-based methods is badly affected by the choice of propensity estimation models and the variance of the estimator. We also show that the proposed method significantly improves the recommendation quality, especially for situations where the propensity score is hard to estimate.

The contributions of this paper can be summarized as follows: 
\begin{itemize}
    \item We propose a model-agnostic meta-learning method. The proposed method is the first method that minimizes the propensity-independent upper bound of the ideal loss function to address the selection bias of recommender systems.
    \item We empirically show that the performance of the propensity-based recommendation is largely affected by choice of propensity estimators and the variance caused by the inverse propensity weighting.
    \item We demonstrate that the proposed method stably improve the recommendation quality, especially when it is difficult to estimate the propensity score.
\end{itemize}
\section{Related Work}
In this section, we review existing related studies.

\subsection{Propensity-based Recommendation}
Propensity-based methods aim to accurately estimate the loss function of interest using only biased rating feedback~\cite{schnabel2016recommendations, liang2016causal, wang2018the}.
The probability of observing each entry of a rating matrix is defined as the propensity score, and the unbiased estimator for the metric of interest can be derived by weighting each sample by the inverse of its propensity~\cite{rosenbaum1983central, rubin1974estimating, imbens2015causal}. The method of matrix factorization with IPS (MF-IPS)~\cite{schnabel2016recommendations,liang2016causal} has been demonstrated to outperform naive matrix factorization and probabilistic generative models ~\cite{hernandez2014probabilistic,mnih2008probabilistic} under MNAR settings. Moreover, the DR estimation, used in the off-policy evaluation of the bandit algorithms~\cite{dudik2011doubly,jiang2016doubly}, has also been applied to the MNAR recommendation~\cite{wang2019doubly}. The DR estimation combines the propensity score estimation and the error imputation model in a theoretically sophisticated manner and improves the statistical properties of the IPS estimator. The error imputation model is the model of predicted errors for the missing ratings, and the performance of the DR estimator has been proven to be dependent on the accuracy of the propensity score estimation and the error imputation model~\cite{wang2019doubly}. 

All the methods stated above are based on explicit feedback. For recommendations using MNAR implicit feedback,~\cite{saito2020unbiased} is the first work to construct an unbiased estimator for the loss function of interest using only biased implicit feedback. The proposed estimator is a combination of the IPS estimation and positive-unlabeled learning~\cite{bekker2018learning,elkan2008learning}.
Although generalization error analysis is not conducted in this work, it is shown that the variance of the propensity-based estimator depends on the inverse of the propensity score, and this leads to a severe variance problem, especially when there exists severe selection bias.

These propensity-based algorithms utilize the unbiased loss function; however, the performance of these methods largely depends on the propensity score estimation. Ensuring the accuracy of the propensity score estimation is difficult for real-world recommenders~\cite{yang2018unbiased}, as the analysts cannot control the missing mechanism. Thus, methods to improve the robustness of propensity-based approaches are highly desired.

\subsection{Off-policy Evaluation and Learning}
Off-policy evaluation aims to accurately evaluate the performance of contextual bandit policies in offline settings~\cite{dudik2011doubly,farajtabar2018more,gilotte2018offline}. Most existing off-policy estimators utilize the Direct Method (DM) or the IPS estimation technique~\cite{dudik2011doubly,rosenbaum1983central,rubin1974estimating,imbens2015causal}. DM predict the reward function using the logged bandit feedback with arbitrary machine learning algorithms, and then, use these predictions to estimate the performance of a given policy. In contrast, the IPS estimation approach uses the propensity score and corrects the distributional shift between the past policy and the new policy that is to be evaluated. It is widely known that the DM approach is subject to the bias problem, and the IPS approach is subject to the variance problem~\cite{dudik2011doubly}. To explore the best bias-variance trade-off for off-policy evaluation, several combinations of the DM and the IPS approach have been proposed, including DR~\cite{dudik2011doubly}, a more robust doubly robust~\cite{farajtabar2018more}, or SWITCH estimator~\cite{wang2017optimal}.

In contrast, off-policy learning aims to obtain a well-performing action policy offline using only logged bandit feedback~\cite{swaminathan2015counterfactual,joachims2018deep}. The fundamental work for developing the off-policy optimization procedure was carried out by~\cite{swaminathan2015counterfactual}; in their paper, they propose the counterfactual risk minimization framework and a corresponding algorithm called POEM. It optimizes the lower bound of the performance of action policies, and this lower bound consists of the mean and variance of the IPS estimator. The other promising approach for off-policy learning is the DACPOL procedure proposed in~\cite{atan2018learning}, where the propensity-independent lower bound is optimized via adversarial learning. The derived lower bound consists of the empirical policy performance based on observational data and a distance measure for the distributional shift between the randomized and observational data. The DACPOL
procedure empirically outperforms the propensity-based POEM algorithm in situations where past treatment policies (propensities) are unknown~\cite{atan2018learning}.

The propensity-based recommendation methods summarized in Section 2.1 are similar to the off-policy evaluation, as both aim to unbiasedly estimate the metric of interest, for example, by using the propensity weighting estimator. However, the upper bound minimization approach, such as the DACPOL framework for off-policy learning, is also theoretically sound and has shown its strength empirically in learning situations when the propensity score is unknown. A method that uses the upper bound minimization approach has not yet been proposed for MNAR recommendation settings, despite that propensity estimation is difficult in real-life recommender systems due to several confounding factors~\cite{joachims2017unbiased,wang2018position,li2018offline}.

\subsection{Unsupervised Domain Adaptation}
Unsupervised domain adaptation aims to train a predictor that works well on a target domain by using only labeled source samples and unlabeled target samples during training~\cite{saito2017asymmetric}. One difficulty is that the feature distributions and the labeling functions\footnote{mapping from feature space to outcome space} are different between the source and target domains. Therefore, a model trained on the source domain does not generalize well to the target domain and measuring the difference between the two domains is critical~\cite{lee2019domain}. Some discrepancy measures to measure this difference have been proposed. Among them, $\mathcal{H}$-divergence and $\mathcal{H}\Delta\mathcal{H}$-divergence~\cite{ben2007analysis, ben2010theory} have been used to construct many prediction methods. For example, the domain adversarial neural network simultaneously minimizes source empirical errors and the $\mathcal{H}$- divergence between the source and target domains in an adversarial manner~\cite{ganin2015unsupervised, ganin2016domain}. The asymmetric tri-training framework trains three networks asymmetrically and is interpreted as minimizing the $\mathcal{H}$-divergence during training~\cite{saito2017asymmetric}.

Our proposed method is based on the upper bound minimization framework, which has shown its effectiveness in unsupervised domain adaptation settings. This work is the first to extend the upper bound minimization approach to MNAR recommendation, and we demonstrate its advantages over the unbiased estimation approach in Section 5.
\section{Preliminaries}
In this section, we introduce the basic notation and formulation of the MNAR explicit recommendation.

\subsection{Problem Formulation}
Let $\mathcal{U}$ be a set of users ($|\mathcal{U}| = m$), and $\mathcal{I}$ be a set of items ($ |\mathcal{I}| = n$).
We denote the set of all user and item pairs as $\mathcal{D} = \mathcal{U} \times \mathcal{I}$. Let $\boldsymbol{R} \in \mathbb{R}^{m \times n}$ be a true rating matrix; each entry $R_{u,i}$ is the true rating of user $u$ to item $i$.

The focus of this study is to establish an algorithm to obtain an optimal predicted rating matrix denoted as $\widehat{\boldsymbol{R}}$.
Each entry $\widehat{R}_{u,i}$ is the predicted rating for the user-item pair $(u, i)$.
To achieve this goal, we formally define the ideal loss function of interest that should be minimized to derive the predictions as
\begin{align}
    \mathcal{L}^{\ell}_{ideal} \left( \widehat{\boldsymbol{R}} \right) = \frac{1}{|\mathcal{D}|} \sum_{(u,i) \in \mathcal{D}} \ell \left(R_{u,i}, \widehat{R}_{u,i} \right) \label{eq:ideal}
\end{align}
where $\ell (\cdot, \cdot) : \mathbb{R} \times \mathbb{R} \rightarrow \mathbb{R}_{\ge 0} $ is an arbitrary loss function.
For example, when $\ell (x, y) = (x - y)^2$, \Eqref{eq:ideal} is the mean-squared-error (MSE).

In reality, it is impossible to calculate the ideal loss function, as most of the true ratings are missing.
To formulate the missing mechanism of the true ratings, we introduce another matrix $ \boldsymbol{O} \in \{ 0, 1\}^{m \times n} $ called the indicator matrix, and each entry $O_{u,i} $ is a Bernoulli random variable representing whether the true rating of $(u, i)$ is observed.
If $O_{u,i} = 1$, then $R_{u,i}$ is observed; otherwise, $R_{u,i}$ is unobserved. Using indicator variables, we can denote the set of user-item pairs for the observed ratings as $ \mathcal{O} = \{ (u,i) \, | \, O_{u,i} = 1 \} $. When the missing mechanism is MNAR, accurately estimating the ideal loss function using the observed dataset $\mathcal{O}$ is essential to derive an effective recommender.

\subsection{Naive Estimator}
The simplest estimator for the ideal loss function is called the naive estimator, which is defined as follows:

\begin{align*}
\widehat{\mathcal{L}}^{\ell}_{naive} \left( \widehat{\boldsymbol{R}} \right)
= \frac{1}{ | \mathcal{O} | } \sum_{(u,i) \in \mathcal{O}} \ell \left( R_{u,i}, \widehat{R}_{u,i} \right)
\end{align*}

This estimator calculates the average loss function over the observed ratings, and most existing methods are based on this simple estimator.
If the missing ratings are MCAR, the naive estimator is unbiased against the ideal loss function. However, in the case of MNAR datasets, the naive estimator is biased \cite{steck2010training, schnabel2016recommendations}, i.e., 
\begin{align}
    \mathbb{E}_{ \boldsymbol{O} } \left[ \widehat{\mathcal{L}}^{\ell}_{naive} \left( \widehat{\boldsymbol{R}} \right) \right]
    \neq \mathcal{L}_{ideal} \left( \widehat{\boldsymbol{R}} \right) \label{eq:naive}
\end{align}
for some given $\widehat{\boldsymbol{R}}$. Thus, one has to use an estimator that can address this bias issue alternative to using the naive one

\subsection{Inverse Propensity Score Estimator}
In \cite{schnabel2016recommendations, liang2016causal}, the authors applied the IPS estimation to address the bias under the MNAR mechanism. The propensity scoring method has been previously proposed in the context of causal inference to estimate treatment effects using observational data \cite{rosenbaum1983central, rubin1974estimating, imbens2015causal}. The basic idea of this estimator is to create a pseudo-MCAR dataset by weighting the observed ratings by the inverse of its propensity score.

In this work, the propensity score of user-item pair $(u, i)$ is formally defined as $P_{u,i} = \mathbb{P} \left( O_{u,i} = 1 \right) = \mathbb{E} \left[ O_{u,i} \right]  $. By using the propensity score, the unbiased estimator for the ideal loss function can be derived as follows:
\begin{align}
\widehat{\mathcal{L}}^{\ell}_{IPS} \left( \widehat{\boldsymbol{R}} \right) 
& = \frac{1}{ | \mathcal{D} | } \sum_{(u,i) \in \mathcal{D}} O_{u,i} \cdot \frac{ \ell \left( R_{u,i}, \widehat{R}_{u,i} \right)} { P_{u,i} } \label{eq:ips}
\end{align}

This estimator is unbiased against the ideal loss function, i.e.,
\begin{align*}
\mathbb{E}_{ \boldsymbol{O} } \left[ \widehat{\mathcal{L}}^{\ell}_{IPS} \left( \widehat{\boldsymbol{R}} \right) \right]
= \mathcal{L}^{\ell}_{ideal} \left( \widehat{\boldsymbol{R}} \right)
\end{align*}
for any given $\widehat{\boldsymbol{R}}$, and thus considered to be more desirable than the naive estimator.

As theoretically and empirically stated in \cite{schnabel2016recommendations}, unbiasedness of the IPS estimator is desirable; however, this property depends on the true propensity score. In reality, the true propensity score is unobservable and thus has to be estimated using the naive Bayes, logistic regression, or Poisson factorization \cite{schnabel2016recommendations, liang2016causal}. If the propensity estimation model is misspecified, the IPS estimator is no longer an unbiased estimator. Moreover, the IPS estimator often suffers from a high variance, as the inverse of the propensities might be large \cite{saito2019doubly,dudik2011doubly}. 

These problems can also be theoretically explained.
\begin{theorem}(Theorem 5.2 of \cite{schnabel2016recommendations}) Suppose that the loss function is bounded above by a positive constant $\Delta$. Then, for any finite hypothesis space of predictions $ \mathcal{H}=\{\widehat{\boldsymbol{R}}_{1}, \ldots, \widehat{\boldsymbol{R}}_{|\mathcal{H}|}\} $ and for any $\delta \in (0,1)$, the following inequality holds with a probability of at least $1 - \delta$.
\begin{align*}
    \mathcal{L}_{ideal} \left(\widehat{\boldsymbol{R}}_{ERM} \right) 
    & \le \widehat{\mathcal{L}}_{IPS}\left(\widehat{\boldsymbol{R}}_{ERM}  \, | \, \boldsymbol{O}\right)  
    + \underbrace{\frac{\Delta}{ |\mathcal{D}| } \sum_{(u, i) \in \mathcal{D}}\left|1-\frac{P_{u, i}}{\hat{P}_{u, i}}\right|}_{{\textit{bias term}}} \\
    & + \underbrace{\frac{\Delta}{|\mathcal{D}|} \sqrt{\frac{1}{2} \log \frac{2|\mathcal{H}|}{\delta} } \sqrt{\sum_{(u, i) \in \mathcal{D}} \frac{1}{\hat{P}^2_{u,i}}}}_{{\textit{variance term}}}
\end{align*}
where $\widehat{P}_{u,i}$ is an estimated value for $P_{u,i}$, and 
$$
    \widehat{\boldsymbol{R}}_{ERM} = \argmin_{\widehat{\boldsymbol{R}} \in \mathcal{H} } \, \widehat{\mathcal{L}}_{IPS} (\widehat{\boldsymbol{R}} \ | \ \boldsymbol{O} )
$$ 
is the empirical risk minimizer.
\label{theorem1}
\end{theorem}

The generalization error bound of the empirical risk minimizer in \Thmref{theorem1} depends on both the bias and variance terms. When the estimation error of the propensity estimator is large, the bias term can also be large. Moreover, the variance term depends on the inverse of the estimated propensity scores; the variance problem results in a loose generalization upper bound.

Therefore, developing learning methods that are robust to the propensity misspecification and the variance of the estimator is critical to apply the methods to real-world MNAR problems.
\section{Method}

\begin{algorithm}[t]
    \caption{Asymmetric tri-training procedure for missing-not-at-random explicit feedback}
    \begin{algorithmic}[1]
        \Require observed rating dataset $\mathcal{O}$, three predictors $A_1, A_2, A_3$, set of hyperparameters \{$\epsilon$, \textit{number of iterations}, \textit{number of steps}\}
        \Ensure predicted rating matrix $ \widehat{\boldsymbol{R}}$ by $A_3$
        \State Pre-train $A_1, A_2, A_3$ using the observed rating dataset $\mathcal{O}$
        \State randomly sample user-item pairs from $\mathcal{D}$ to generate $\mathcal{D}'$
        \State generate a dataset with pseudo-ratings $\widetilde{\mathcal{D}}$ by \Eqref{eq:pseudo_dataset}
        \For {$i=1$ to \textit{number of iterations}}
        \For {$j=1$ to \textit{number of steps}}
        \State Update $A_1$ and $A_2$ with mini-batch data from $\widetilde{\mathcal{D}}$
        \State Update $A_3$ with mini-batch data from $\widetilde{\mathcal{D}}$
        \EndFor
        \State randomly sample user-item pairs from $\mathcal{D}$ to generate $\mathcal{D}'$
        \State generate a dataset with pseudo-ratings $\widetilde{\mathcal{D}}$ by \Eqref{eq:pseudo_dataset}
        \EndFor
        \State \Return $ \widehat{\boldsymbol{R}}$ by $A_3$
    \end{algorithmic}
\end{algorithm}

\subsection{Meta-learning procedure}
To realize the objective with only biased rating feedback, we propose the \textit{asymmetric tri-training} framework that utilizes three rating predictors asymmetrically. First, two of the three predictors are trained to generate a reliable dataset with pseudo ratings.
Then, the other predictor is trained on that pseudo-ratings. 
We can use any recommendation algorithm, such as matrix factorization \cite{mnih2008probabilistic, koren2009matrix}, MF-IPS \cite{schnabel2016recommendations, liang2016causal}, factorization machines \cite{rendle2010factorization}, and neural network matrix factorization \cite{dziugaite2015neural}, for the three predictors. 
Thus, the proposed method is highly general and can be used to improve the prediction accuracy of methods proposed in the future.

The asymmetric tri-training framework consists of three steps.
First, in the \textbf{pre-training step}, we pre-train the three selected recommendation algorithms $A_1, A_2$, and $A_3$ using the observed rating data $\mathcal{O}$. Next, we randomly sample user-item pairs\footnote{This sampling is optional; one can simply use $\mathcal{D}$ as the dataset $\mathcal{D}'$.}, denoted as $ \mathcal{D}'$. Then, we predict the ratings of the unlabeled dataset $\mathcal{D}'$ using two of the three algorithms $A_1$ and $A_2$. The predicted rating for $(u, i) \in \mathcal{D}'$ by $A_1$ and $A_2$ is denoted as $ \widehat{R}_{u,i}^{(1)} $ and $ \widehat{R}_{u,i}^{(2)} $, respectively. 
We regard one of the two predicted values as the pseudo-rating for $(u, i)$ if the two predicted values are sufficiently similar. 
By doing this, we can construct a dataset with reliable pseudo-ratings.
The resulting dataset is denoted as
\begin{align}
    \widetilde{\mathcal{D}} = \left\{ \bigl(u, i, \widehat{R}^{(1)}_{u,i} \bigr) : (u,i) \in \mathcal{D}^{\prime}, \bigl| \widehat{R}_{u,i}^{(1)}  - \widehat{R}_{u,i}^{(2)} \bigr| \le \epsilon \right\} \label{eq:pseudo_dataset}
\end{align}
where $\epsilon > 0$ is a hyperparameter and should be tuned via a parameter tuning procedure. This step is the \textbf{pseudo labeling step}.

Finally, we train the remaining predictor $A_3$ by minimizing the following loss function.
\begin{align*}
\widehat{\mathcal{L}}^{\ell}_{pseudo} \left( \widehat{\boldsymbol{R}} , \widehat{\boldsymbol{R}}^{(1)}   \right)
& = \frac{1}{ | \widetilde{\mathcal{D}} | } \sum_{ (u,i) \in \widetilde{\mathcal{D}} } \ell \left(\widehat{R}_{u,i} , \widehat{R}_{u,i}^{(1)}   \right) \notag \\
& = \frac{1}{ | \widetilde{\mathcal{D}} | } \sum_{ (u,i) \in \mathcal{D}} O^{\prime}_{u,i} \cdot \ell \left(\widehat{R}_{u,i} , \widehat{R}_{u,i}^{(1)}   \right)
\end{align*}
where $ \{  \widehat{R}_{u,i}^{(1)} \} $ are the pseudo-ratings provided by $A_1$ , $ \{  \widehat{R}_{u,i} \} $ are the predicted ratings provided by $A_3$, and $ \{ O^{\prime}_{u,i} \}$ are other indicator variables representing whether the user-item pairs are in the created pseudo-labeled dataset $\widetilde{\mathcal{D}}$. This step is called the \textbf{final prediction step}.

In the algorithm, we iterate the pseudo-labeling step several times to generate a reliable pseudo ratings.
Algorithm 1 describes the complete learning procedure of the asymmetric tri-training.

\subsection{Theoretical Analysis}
In this subsection, we theoretically analyze the MNAR recommendation. 
Specifically, we drive the propensity-independent upper bound of the ideal loss function and demonstrate that the proposed asymmetric tri-training framework attempts to minimize the part of the upper bound while keeping it informative during training.

In the following proposition, we first derive a simple upper bound of the ideal loss function based on the triangle inequality.
\begin{proposition} Suppose that the loss function $\ell$ obeys the triangle inequality. Then, for any given predicted rating matrices $ \widehat{ \boldsymbol{R}}^{(1)} $, $ \widehat{ \boldsymbol{R}}^{(2)} $, and $ \widehat{ \boldsymbol{R}} $, the following inequality holds.
    \begin{align*}
    \mathcal{L}^{\ell}_{ideal} \left(\widehat{\boldsymbol{R}}, \boldsymbol{R} \right)
    & \le \mathcal{L}^{\ell}_{ideal} \left( \widehat{\boldsymbol{R}}, \widehat{\boldsymbol{R}}^{(1)} \right) \notag \\
    & + \mathcal{L}^{\ell}_{ideal} \left( \widehat{\boldsymbol{R}}^{(1)}, \widehat{\boldsymbol{R}}^{(2)} \right)
    + \mathcal{L}^{\ell}_{ideal} \left( \widehat{\boldsymbol{R}}^{(2)}, \boldsymbol{R} \right)
    \end{align*}
    \begin{proof}
        We apply the triangle inequality twice:
        \begin{align}
        & \mathcal{L}^{\ell}_{ideal} \left( \widehat{\boldsymbol{R}}, \boldsymbol{R} \right) \notag \\
        & \le \mathcal{L}^{\ell}_{ideal} \left( \widehat{\boldsymbol{R}}, \widehat{\boldsymbol{R}}^{(1)} \right) + \mathcal{L}^{\ell}_{ideal} \left( \widehat{\boldsymbol{R}}^{(1)}, \boldsymbol{R} \right)  \notag \\
        & \le \mathcal{L}^{\ell}_{ideal} \left( \widehat{\boldsymbol{R}}, \widehat{\boldsymbol{R}}^{(1)} \right) +  \mathcal{L}^{\ell}_{ideal} \left( \widehat{\boldsymbol{R}}^{(1)}, \widehat{\boldsymbol{R}}^{(2)} \right) +
        \mathcal{L}^{\ell}_{ideal} \left( \widehat{\boldsymbol{R}}^{(2)}, \boldsymbol{R} \right)    \notag
        \end{align}
    \end{proof}
    \label{proposition1}
\end{proposition}

We further analyze the propensity-independent upper bound of the ideal loss function.

\begin{lemma}(Hoeffding's Inequality)
    Independent bounded random variables $Z_1 , ..., Z_n$ that take values in intervals of sizes $\zeta_1 , ..., \zeta_n$ satisfy the following inequality for any $\eta > 0$.
    \begin{align*}
    \mathbb{P} \left( \left|  \sum_{i=1}^n Z_i - \mathbb{E} \left[ \sum_{i=1}^n Z_i \right] \right| \geq \eta  \right) \leq 2 \exp \left( \frac{-2 \eta^2}{ \sum_{i=1}^n \zeta_i^2 } \right) 
    \end{align*}
    \label{lem1}
    See Theorem 2 in \cite{hoeffding1994probability} for the proof.
    \label{lemma1}
\end{lemma}

\begin{theorem}(Propensity-independent generalization error bound)
Suppose that a pseudo-labeled dataset $\widetilde{D}$, and two predicted matrices $ \widehat{\boldsymbol{R}}^{(1)}$ and $\widehat{\boldsymbol{R}}^{(2)} $ are given. In addition, a loss function $\ell$ obeys the triangle inequality and is bounded above by a positive constant $\Delta$.
Then, for any $\widehat{\boldsymbol{R}} \in \mathcal{H}$, where $\mathcal{H} = \{ \widehat{\boldsymbol{R}}_1, \ldots, \widehat{\boldsymbol{R}}_{ | \mathcal{H}| } \} $ is a given finite hypothesis space, and for any $\delta \in (0, 1)$, the following inequality holds with a probability of at least $1 - \delta$.
\begin{align*}
& \mathcal{L}^{\ell}_{ideal} \left( \widehat{\boldsymbol{R}}, \boldsymbol{R} \right) \\
& \le \underbrace{\widehat{\mathcal{L}}^{\ell}_{pseudo} \left( \widehat{\boldsymbol{R}}, \widehat{\boldsymbol{R}}^{(1)} \right)}_{(a)} 
+ bias \left(\widehat{\mathcal{L}}^{\ell}_{pseudo} \left( \widehat{\boldsymbol{R}}, \widehat{\boldsymbol{R}}^{(1)} \right) \right) \notag \\
& + \underbrace{\mathcal{L}^{\ell}_{ideal} \left( \widehat{\boldsymbol{R}}^{(1)}, \widehat{\boldsymbol{R}}^{(2)} \right)}_{(b)} 
+ \underbrace{\mathcal{L}^{\ell}_{ideal} \left( \widehat{\boldsymbol{R}}^{(2)}, \boldsymbol{R} \right)}_{(c)} 
+ \frac{\Delta}{|\widetilde{\mathcal{D}}|} \sqrt{  \frac{ |\mathcal{D}| }{2 } \log \left( \frac{2 \left| \mathcal{H} \right| }{ \delta} \right) }
\end{align*}
where
\begin{align}
bias \left(\widehat{\mathcal{L}}^{\ell}_{pseudo} \left( \widehat{\boldsymbol{R}}, \widehat{\boldsymbol{R}}^{(1)} \right) \right)
& = \mathcal{L}^{\ell}_{ideal} \left( \widehat{\boldsymbol{R}}, \widehat{\boldsymbol{R}}^{(1)} \right) - \mathbb{E} \left[ \widehat{\mathcal{L}}^{\ell}_{pseudo} \left( \widehat{\boldsymbol{R}}, \widehat{\boldsymbol{R}}^{(1)} \right) \right] \notag
\end{align}
\begin{proof}
    We prove that the following inequality holds with a probability of at least $1 - \delta$:
    \begin{align}
    \mathcal{L}^{\ell}_{ideal} \left( \widehat{\boldsymbol{R}}, \widehat{\boldsymbol{R}}^{(1)} \right)
    & \le \widehat{\mathcal{L}}^{\ell}_{pseudo} \left( \widehat{\boldsymbol{R}}, \widehat{\boldsymbol{R}}^{(1)} \right) \notag \\
    & + bias \left(\widehat{\mathcal{L}}^{\ell}_{pseudo} \left( \widehat{\boldsymbol{R}}, \widehat{\boldsymbol{R}}^{(1)} \right) \right)
    + \frac{\Delta}{|\widetilde{\mathcal{D}}|} \sqrt{  \frac{ |\mathcal{D}| }{2 } \log \left( \frac{2 \left| \mathcal{H} \right| }{ \delta} \right) } \notag \\ \label{eq:final_bound}
    \end{align}
    First, the following equation holds:
    \begin{align}
    & \mathcal{L}^{\ell}_{ideal} \left( \widehat{\boldsymbol{R}}, \widehat{\boldsymbol{R}}^{(1)} \right) \notag \\
    & = \mathcal{L}^{\ell}_{ideal} \left( \widehat{\boldsymbol{R}}, \widehat{\boldsymbol{R}}^{(1)} \right)
    - \mathbb{E} \left[ \widehat{\mathcal{L}}^{\ell}_{pseudo} \left( \widehat{\boldsymbol{R}}, \widehat{\boldsymbol{R}}^{(1)} \right) \right]
    + \mathbb{E} \left[ \widehat{\mathcal{L}}^{\ell}_{pseudo} \left( \widehat{\boldsymbol{R}}, \widehat{\boldsymbol{R}}^{(1)} \right) \right] \notag \\
    & = \mathbb{E} \left[ \widehat{\mathcal{L}}^{\ell}_{pseudo} \left( \widehat{\boldsymbol{R}}, \widehat{\boldsymbol{R}}^{(1)} \right) \right]
    + bias \left( \widehat{\mathcal{L}}^{\ell}_{pseudo} \left( \widehat{\boldsymbol{R}}, \widehat{\boldsymbol{R}}^{(1)} \right)  \right)  \label{eq:bias}
    \end{align}
    Here, $\{ O^{\prime}_{u,i} \}$ are independent from assumption, and we apply Hoeffding's inequality in \Lemref{lemma1} to $ \widehat{\mathcal{L}}^{\ell}_{pseudo} ( \widehat{\boldsymbol{R}}, \widehat{\boldsymbol{R}}^{(1)} ) $, which yields:
    \begin{align}
    & \mathbb{P} \left( \left| \mathbb{E} \left[ \widehat{\mathcal{L}}^{\ell}_{pseudo} \left( \widehat{\boldsymbol{R}}, \widehat{\boldsymbol{R}}^{(1)} \right) \right]  - \widehat{\mathcal{L}}^{\ell}_{pseudo} \left( \widehat{\boldsymbol{R}}, \widehat{\boldsymbol{R}}^{(1)} \right) \right| \ge \eta \right) \notag \\
    & \le \mathbb{P} \left( \max_{ \widehat{\boldsymbol{R}}' \in \mathcal{H} } \left| \mathbb{E} \left[ \widehat{\mathcal{L}}^{\ell}_{pseudo} \left( \widehat{\boldsymbol{R}}^{\prime}, \widehat{\boldsymbol{R}}^{(1)} \right) \right]  - \widehat{\mathcal{L}}^{\ell}_{pseudo} \left( \widehat{\boldsymbol{R}}^{\prime}, \widehat{\boldsymbol{R}}^{(1)} \right) \right| \ge \eta \right) \notag \\
    & \le  \mathbb{P} \left( \bigvee_{ \widehat{\boldsymbol{R}}' \in \mathcal{H} } \left| \mathbb{E} \left[ \widehat{\mathcal{L}}^{\ell}_{pseudo} \left( \widehat{\boldsymbol{R}}^{\prime}, \widehat{\boldsymbol{R}}^{(1)} \right) \right]  - \widehat{\mathcal{L}}^{\ell}_{pseudo} \left( \widehat{\boldsymbol{R}}^{\prime}, \widehat{\boldsymbol{R}}^{(1)} \right) \right| \ge \eta \right) \notag \\
    & \le \sum_{ \widehat{\boldsymbol{R}}' \in \mathcal{H} } 2 \exp \left( \frac{-2 | \mathcal{\widetilde{D}} |^2  \eta^2}{ |\mathcal{D}| \Delta^2  }  \right)
    \le 2 \left| \mathcal{H} \right| \exp \left( \frac{-2 | \mathcal{\widetilde{D}} |^2  \eta^2}{ |\mathcal{D}| \Delta^2  }  \right) \notag
    \end{align}
    We set $ \delta =  2 \left| \mathcal{H} \right| \exp \left( \frac{-2 | \mathcal{\widetilde{D}} |^2  \eta^2}{ |\mathcal{D}| \Delta^2  }  \right) $, and solving it for $\eta$ yields:
    \begin{align}
    & \mathbb{P} \left( \left| \mathbb{E} \left[ \widehat{\mathcal{L}}^{\ell}_{pseudo} \left( \widehat{\boldsymbol{R}}, \widehat{\boldsymbol{R}}^{(1)} \right) \right]  - \widehat{\mathcal{L}}^{\ell}_{pseudo} \left( \widehat{\boldsymbol{R}}, \widehat{\boldsymbol{R}}^{(1)} \right) \right| \le  t \right) \ge 1 - \delta  \label{eq:hoeffding_bound_2}
    \end{align}
    where 
    $$
    t = \frac{\Delta}{|\widetilde{\mathcal{D}}|} \sqrt{  \frac{ |\mathcal{D}| }{2 } \log \left( \frac{2 \left| \mathcal{H} \right| }{ \delta} \right) }
    $$.
    By combining \Eqref{eq:bias} and \Eqref{eq:hoeffding_bound_2}, \Eqref{eq:final_bound} is obtained. Finally, combining \Propref{proposition1} and \Eqref{eq:final_bound} completes the proof.
\end{proof}
\label{theorem2}
\end{theorem}

\begin{figure*}[ht]
\centering
\begin{center}
    \begin{tabular}{c}
        \begin{minipage}{0.48\hsize}
            \begin{center}
                (a) Yahoo! R3 ($\textit{KL-div}=0.470$)
                \includegraphics[clip, width=8cm]{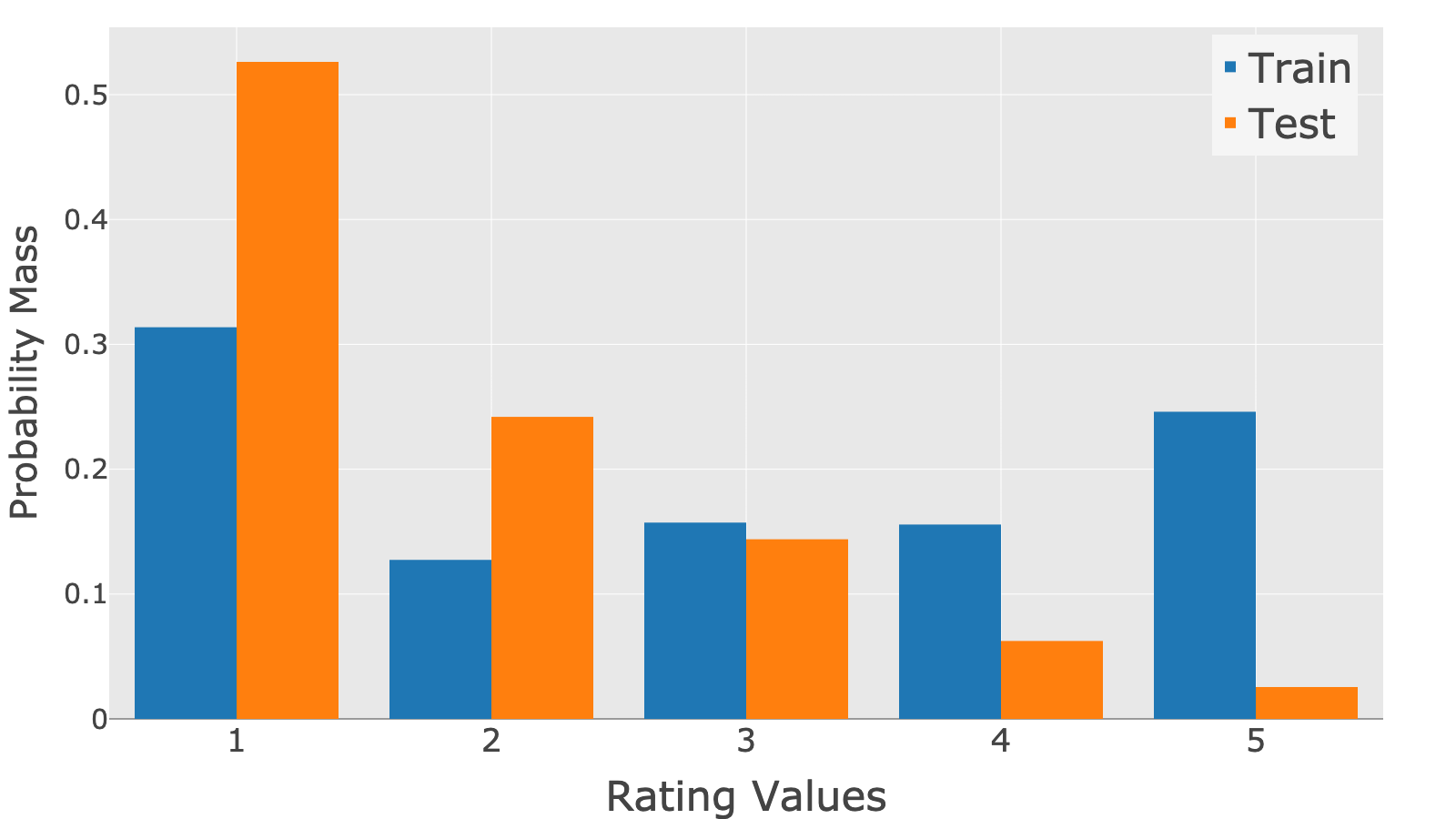}
            \end{center}
        \end{minipage}
        
        \begin{minipage}{0.48\hsize}
            \begin{center}
                (b) Coat ($\textit{KL-div}=0.049$)
                \includegraphics[clip, width=8cm]{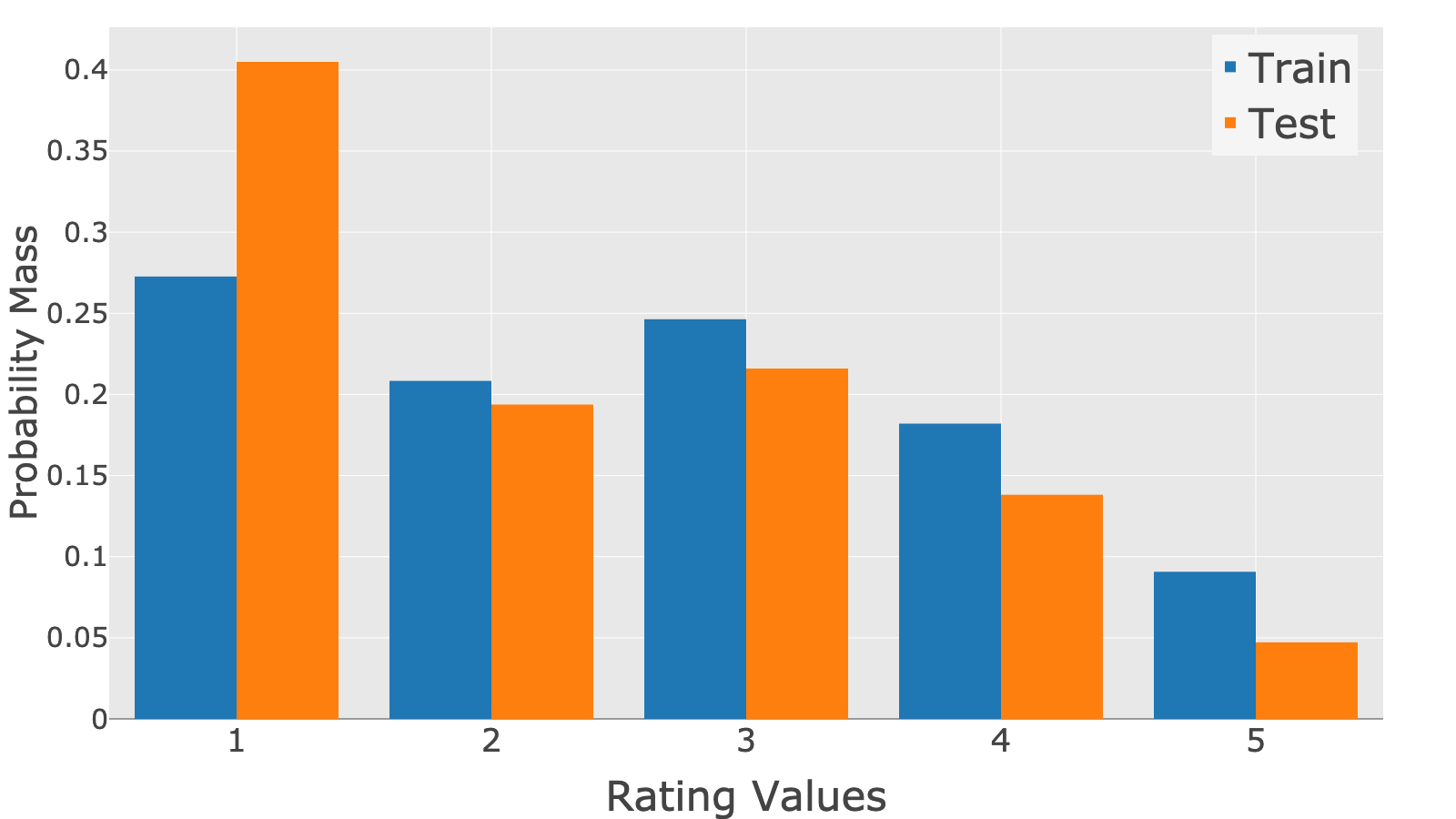}
            \end{center}
        \end{minipage}
    \end{tabular}
\end{center}
\caption{Comparing rating distributions of training and test sets for Yahoo! R3 and Coat datasets}
\vskip 0.05in
\raggedright
\fontsize{9pt}{10pt}\selectfont \textit{Notes}: 
The rating distributions are significantly different between the training and test sets for both datasets. 
Note that \textit{KL-div} is the Kullback–Leibler divergence of the rating distributions between training and test sets.
Therefore, the distributional shift of Yahoo! R3 dataset is relatively large compared to that of the Coat dataset.
\label{fig:rate_dist}
\end{figure*}

As suggested in \Thmref{theorem2}, the following three factors are essential to achieve a small ideal loss: 
\begin{itemize}
    \item[(a)] the loss with respect to the pseudo-ratings.
    \item[(b)] the similarity of the predicted values by $A_1$ and $A_2$.
    \item[(c)] the ideal loss of $A_2$ with respect to the true ratings.
\end{itemize}
Note that the derived upper bound is independent of the propensity score, even if we use propensity-based algorithms for $A_1$ or $A_2$. This is because the pseudo-labeling step and the final prediction step of the proposed learning procedure do not use the propensity scoring technique, and thus, the high variance and the propensity misspecification problems are avoided in our theoretical bound.

It should also be noted that the asymmetric tri-training framework is interpreted as a method that attempts to minimize the part of the upper bound of the ideal loss in \Thmref{theorem2}. 
As described in Algorithm 1, two of the three predictors $A_1$ and $A_2$ are trained independently using the observed rating dataset $\mathcal{O}$. 
Subsequently, the other predictor $A_3$ is trained using the dataset generated by pre-trained $A_1$ and $A_2$. 
In the pseudo-labeling step, $A_1$ and $A_2$ are repeatedly updated with same pseudo-ratings, and expected to be similar as the iterations progress.
Thus, the value of $ (b) $ in the RHS of the upper bound is kept small during the pseudo-labeling step (not minimized), which makes the upper bound informative during training.
In addition, the other predictor $A_3$ is trained using $\widetilde{\mathcal{D}}$, and this minimizes the value of $ (a)$. 
Note that the value of $(c)$ depends on the performance of $A_2$, and thus, the upper bound can be loose when $A_2$ performs poorly. 
Nonetheless, in the experiments, we empirically demonstrate that our method actually minimizes the sum of two terms in the bound (i.e., $(a) + (b)$). 
We also show that minimizing the upper bound of the ideal loss function is an effective approach to further improve the recommendation quality on the test set.
\section{Experimental Results}
We conducted comprehensive experiments using benchmark real-world datasets. 
The code for reproducing the results can be found at \href{https://github.com/usaito/asymmetric-tri-rec-real}{\textbf{\textit{https://github.com/usaito/asymmetric-tri-rec-real}}.}

\subsection{Experimental Setup}
\subsubsection{Datasets and Preprocessing}
We used the following real-world datasets.
\begin{itemize}
    \item MovieLens (ML) 100K dataset\footnote{\href{http://grouplens.org/datasets/movielens/}{http://grouplens.org/datasets/movielens/}}: It contains five-star movie ratings collected from a movie recommendation service, and the ratings are MNAR. This dataset involves approximately 100,000 ratings from 943 users and 1,682 movies. In the experiments, we kept movies that had been rated by at least \textbf{\textit{min\_items}} users, and the values of \textbf{\textit{min\_items}} varied with respect to the experimental settings.
    \item Yahoo! R3 dataset\footnote{\href{http://webscope.sandbox.yahoo.com/}{http://webscope.sandbox.yahoo.com/}}: It contains five-star user-song ratings. The training set consists of approximately 300,000 MNAR ratings of 1,000 songs from 15,400 users, and the test set is collected by asking a subset of 5,400 users to rate ten randomly selected songs. Thus, the test set is regarded as an MCAR dataset.
    \item Coat dataset\footnote{\href{https://www.cs.cornell.edu/~schnabts/mnar/}{https://www.cs.cornell.edu/~schnabts/mnar/}}: It contains five-star user-coat ratings from 290 Amazon Mechanical Turk workers on an inventory of 300 coats. The training set contains 6,500 MNAR ratings collected through self-selections by the Turk workers. In contrast, the test set is MCAR collected by asking the Turk workers to rate 16 randomly selected coats.
\end{itemize}

For the ML 100K dataset, we created a test set with a different \textbf{item distribution} from the original one. We created it by first sampling a test set with 50\% of the original dataset, and then, resampling data from the test set based on the \textbf{inverse} of the relative item probabilities in \Eqref{eq:ml_pscore}. This creates a test set, such that each item has a uniform observed probability.
\begin{align}
P_{*,i} =  \frac{ \sum_{u \in \mathcal{U}}  O_{u, i} } { \max_{i \in \mathcal{I}} \sum_{u \in \mathcal{U}}   O_{u, i} } \label{eq:ml_pscore}
\end{align}

For the Yahoo! R3 and Coat datasets, the original datasets were divided into training and test sets. We randomly selected 10\% of the original training set for the validation set. \Figref{fig:rate_dist} shows the rating distributions of training and test sets for the Yahoo! R3 and Coat datasets. The rating distributions are completely different between the training and test sets, which introduces a severe bias when training a recommendation algorithm.

\begin{figure}[ht]
\centering
\begin{center}
    \begin{tabular}{c}
        \includegraphics[clip, width=8cm]{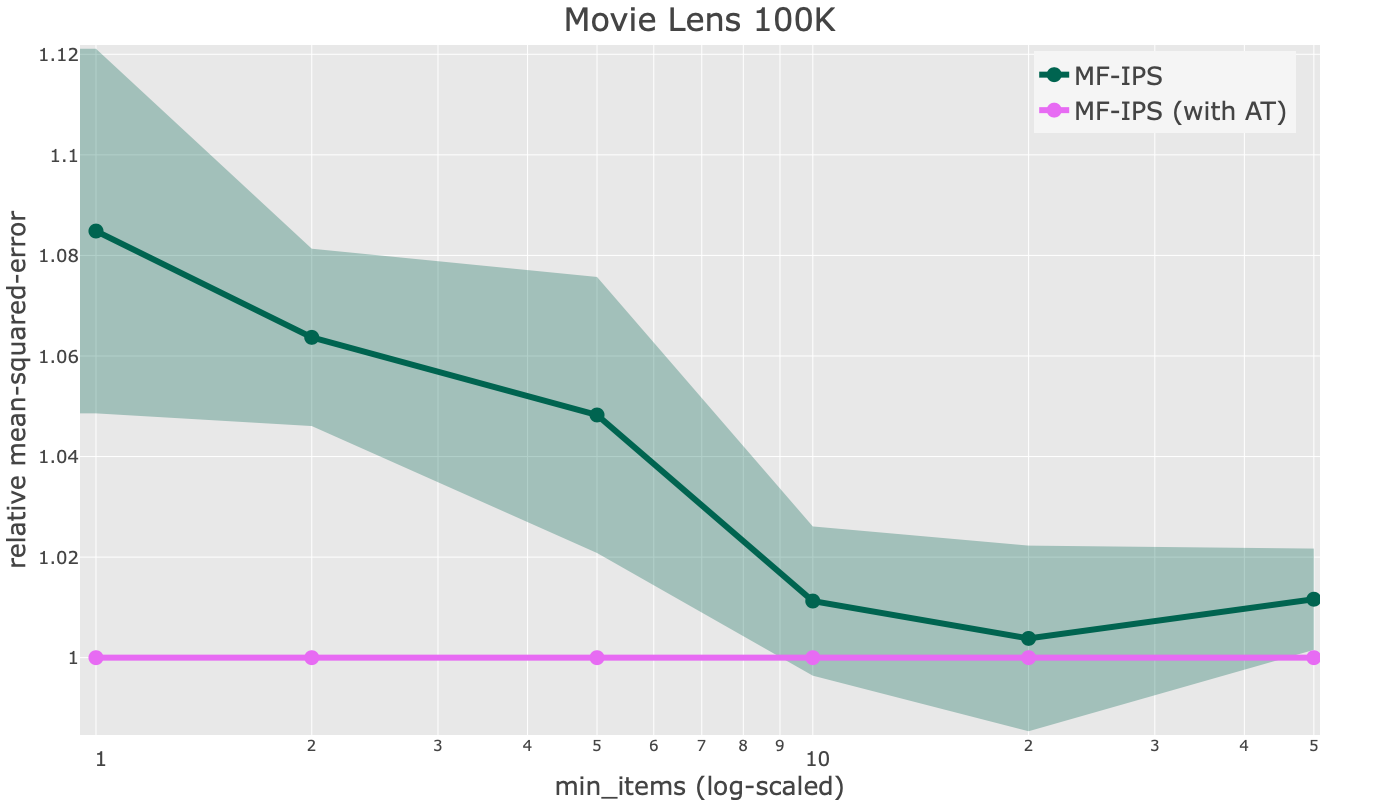}
    \end{tabular}
\end{center}
\caption{Comparing robustness to the variance issue of recommenders with and without \textit{asymmetric tri-training} (AT)}
\vskip 0.05in
\raggedright
\fontsize{9pt}{10pt}\selectfont \textit{Notes}:
The figure reports relative prediction accuracies and their standard errors of MF-IPS with and without AT on a different value of \textit{min\_items}. Both methods were trained with the specified propensity model. MF-IPS with AT significantly outperforms that without AT, especially when a large skewness of the propensity score distribution is present (with a small value of \textit{min\_items}).
\label{fig:variance}
\end{figure}

\subsubsection{Compared methods and propensity estimators}
Here, we describe the baselines and the proposed methods compared in the experiments. We implemented all methods in the \textit{Tensorflow} environment.

\vspace{1mm}
\noindent
\textbf{Matrix Factorization with Inverse Propensity Score (MF-IPS)}:
MF-IPS is based on the MF model \cite{koren2009matrix}. It predicts each rating by $\widehat{R}_{u,i} = \theta_u^{\top} \beta_i + b_u + b_i + b $, where $\{ \theta_u \}$ and $\{ \beta_i \}$ are user and item latent factors, respectively. $b_u$ and $b_i$ are the user and item bias terms. $b$ is the global bias. It optimizes its parameters by minimizing the IPS loss in \Eqref{eq:ips} with regularization terms.

\vspace{1mm}
\noindent
\textbf{MF-IPS with asymmetric-tri training (MF-IPS with AT)}: 
We used MF-IPS with different initializations for $A_1$ and $A_2$ and the MF with naive loss in \Eqref{eq:naive} for $A_3$. 
Thus, the final training step is guaranteed to be independent of the propensity score. \\

For both the baseline and the proposed methods, we tested the following propensity estimators (NB represents \textit{naive Bayes}). 
\begin{align*}
    \textit{uniform propensity} \ : \ & \widehat{P}_{*, *} = \frac{ \sum_{u,i \in \mathcal{D}} O_{u,i} }{ | \mathcal{D} | } \\
    \textit{user propensity} \ : \ & \widehat{P}_{u, *} = \frac{\sum_{i \in \mathcal{I}} O_{u,i}}{\max_{u \in U} \sum_{i \in \mathcal{I}} O_{u,i}} \\
    \textit{item propensity} \ : \ & \widehat{P}_{*, i} = \frac{\sum_{u \in \mathcal{U}} O_{u,i}}{\max_{i \in I} \sum_{u \in \mathcal{U}} O_{u,i}} \\
    \textit{user-item propensity} \ : \ & \widehat{P}_{u, i} = \widehat{P}_{u, *} \cdot \widehat{P}_{*, i} \\
    \textit{NB (uniform)} \ : \ & \widehat{P}_{u,i} =  \mathbb{P}(R=R_{u,i} \, |\, O=1) \mathbb{P}(O=1) \\
    \textit{NB (true)} \ : \ & \widehat{P}_{u,i} = \frac{ \mathbb{P}(R=R_{u,i} \, |\, O=1) \mathbb{P}(O=1) }{ \mathbb{P}(R=R_{u,i}) }
\end{align*}
where $R_{u,i} \in \{1, 2, 3, 4, 5\}$ is a realized rating for $(u,i)$.
Note that when uniform propensity is used, the MF-IPS is identical to the MF with the naive loss function \cite{koren2009matrix}. 
NB (true) is often used as a propensity in previous works \cite{schnabel2016recommendations,wang2019doubly}. 
However, this estimator cannot be used in most real-world problems, as it requires the MCAR explicit feedback to estimate the prior rating distribution (the denominator); we report the results with the this propensity estimator, just for reference.

\begin{table*}[h]
\caption{Comparing prediction and ranking performances of recommenders with and without \textit{asymmetric tri-training} (AT)}
\centering
\scalebox{1.2}{
\begin{tabular}{ccccccccccc} 
\toprule
& && \multicolumn{8}{c}{Metrics} \\ \cmidrule{4-11}
& && \multicolumn{2}{c}{MAE} && \multicolumn{2}{c}{MSE} && \multicolumn{2}{c}{nDCG@3} \\ \cmidrule{4-5}\cmidrule{7-8}\cmidrule{10-11}
Datasets & Propensity && without AT & with AT && without AT & with AT && without AT & with AT \\ \midrule
\multirow{6}{*}{Yahoo! R3} 
& uniform && 1.133& \textbf{0.981} && 1.907 & \textbf{1.452} && 0.351 & \textbf{0.352} \\
& user && 1.062 & \textbf{0.945} && 1.712 & \textbf{1.350} && 0.3523 & \textbf{0.3525} \\ 
 & item && 1.142 & \textbf{0.978} && 1.940 & \textbf{1.458} && 0.351 & \textbf{0.353 } \\ 
& user-item && 1.162 & \textbf{0.991} && 1.979 & \textbf{1.513} && 0.349 & \textbf{0.353}  \\ 
& NB (uniform) && 1.170 & \textbf{1.010} && 1.954 & \textbf{1.511} && 0.351 & \textbf{0.352} \\ \cmidrule{2-11}
& NB (true) && 0.797 & \textbf{0.765} && 1.055 & \textbf{1.014} && 0.351& \textbf{0.353} \\ \midrule 
\multirow{6}{*}{Coat} 
& uniform && \textbf{0.873 }&0.878 && \textbf{1.109} & 1.183 && 0.291 & \textbf{0.293}  \\ 
& user && 0.873 & \textbf{0.832} && \textbf{1.109} & 1.115 && 0.291 & \textbf{0.292}  \\
& item && 0.873 & \textbf{0.832} && 1.117 & \textbf{1.115} && 0.291 & \textbf{0.293} \\
& user-item && 0.874 & \textbf{0.832} && 1.117 & \textbf{1.116} && 0.291 & \textbf{0.293} \\
& NB (uniform) && 0.951 & \textbf{0.920} && 1.268 & \textbf{1.260} && 0.281 & \textbf{0.289} \\ \cmidrule{2-11}
& NB (true) && 0.852 & \textbf{0.831} && \textbf{1.105} & 1.121 && 0.284 & \textbf{0.290} \\ 
\bottomrule
\end{tabular}}
\vskip 0.05in
\raggedright
\fontsize{9pt}{10pt}\selectfont \textit{Notes}: 
For all methods, the average results over 20 different initializations and train-validation splits are reported. 
The results show that the performance of MF-IPS without AT is severely affected by the choice of propensity estimator. 
The proposed method generally improves the rating prediction (MSE and MAE) and ranking quality (nDCG@3) for both datasets. 
Moreover, it demonstrates the robustness to the choice of propensity estimators, especially for Yahoo! R3 data.
\label{tab:results}
\end{table*}

\subsubsection{Hyperparameter Tuning}
For all baselines, the tuning of the L2-regularization hyperparameter was performed in the range of $[10^{-6}, 1]$, and that of the dimensions of the latent factors was performed in the range of $\{5, 10, \ldots, 50\}$. 
For the proposed method, we used the same hyperparameter tuning procedure as with the baselines for the base algorithms ($A_1, A_2, A_3$) and tuned $\epsilon$ in the range of $[10^{-3}, 1]$. 
We searched for an optimal set of hyperparameters using an adaptive procedure implemented in \textit{Optuna} \cite{akiba2019optuna}.
For all methods, we conducted mini-batch optimization with a batch size of $2^{10}$ using the \textit{Adam} optimizer \cite{kingma2014adam} with an initial learning  rate of $0.01$. For the proposed method, we set $\textit{number of iterations}=10$ and $\textit{number of steps}=10$ (see Algorithm 1).
    
\begin{figure*}[t]
\begin{center}
    \begin{tabular}{c}
        \begin{minipage}{0.48\hsize}
            \begin{center}
                (a) Yahoo! R3
                \includegraphics[clip, width=8.5cm]{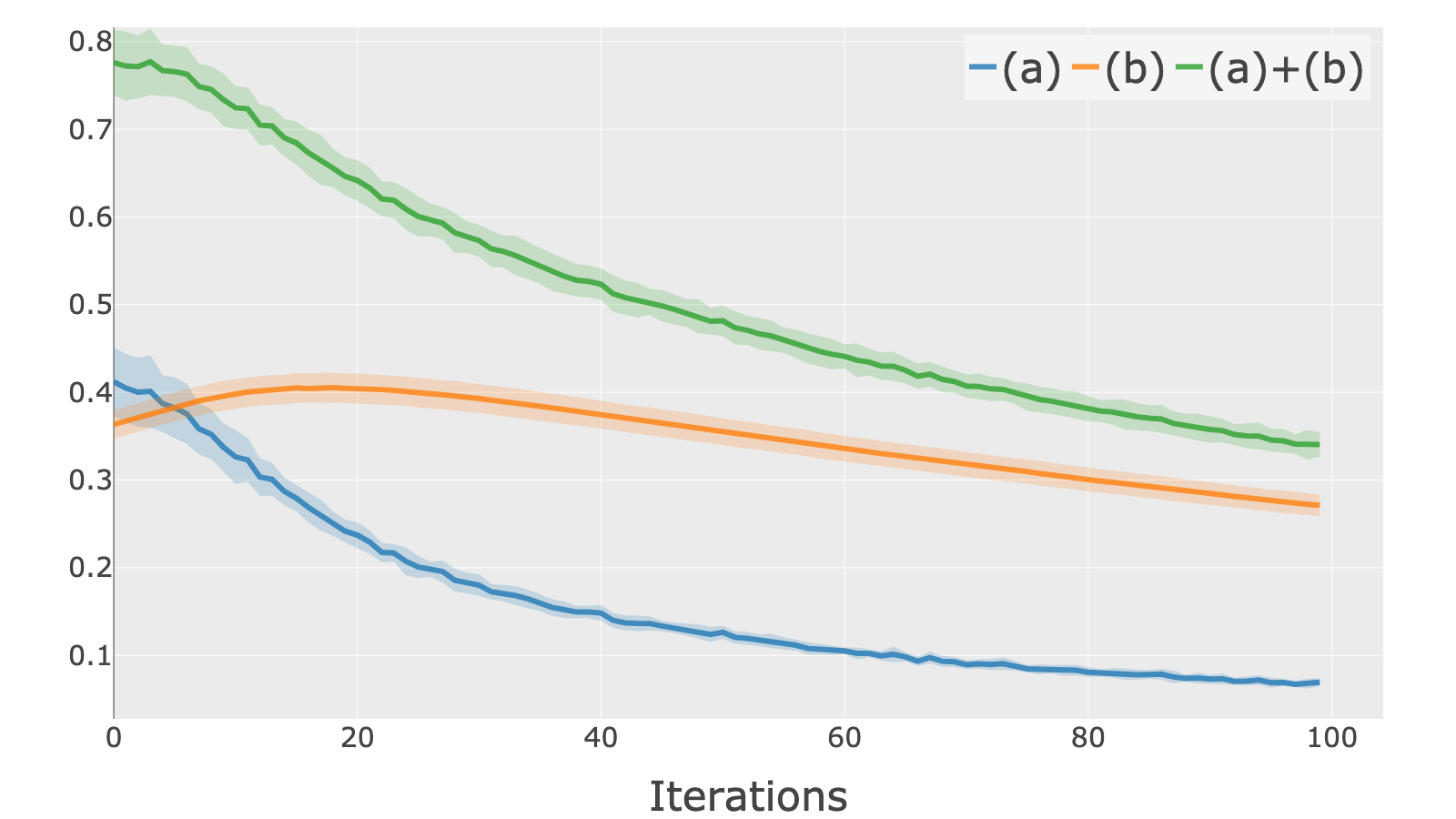}
            \end{center}
        \end{minipage}
        
        \begin{minipage}{0.48\hsize}
            \begin{center}
                (b) Coat
                \includegraphics[clip, width=8.5cm]{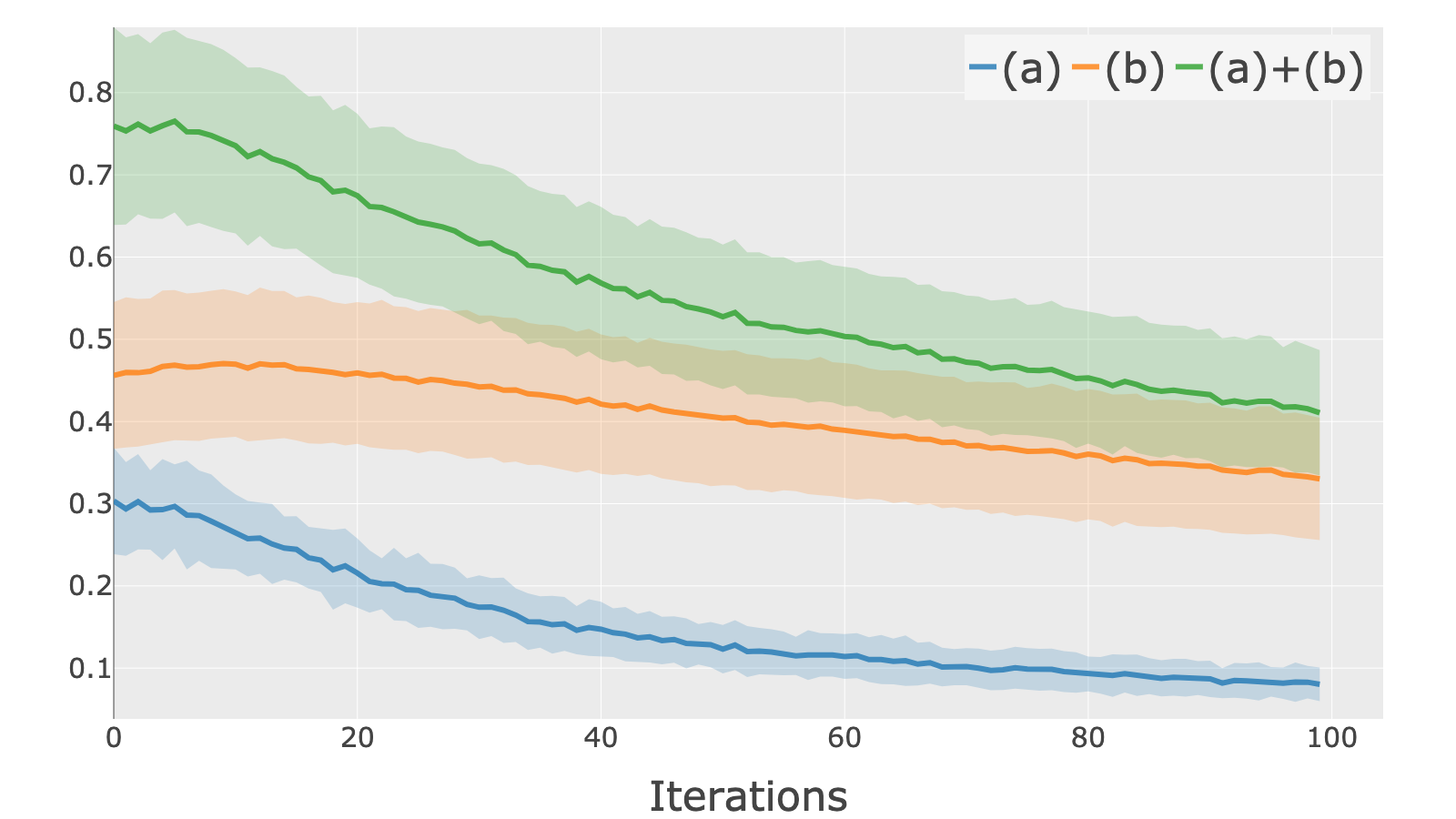}
            \end{center}
        \end{minipage} 
    \end{tabular}
\end{center}
\caption{Upper bound minimization performance of \textit{asymmetric tri-training}}
\vskip 0.05in
\raggedright
\fontsize{9pt}{10pt}\selectfont \textit{Notes}: 
This figure presents averaged values of the loss on generated pseudo ratings ($a$), and the similarity between $A_1$ and $A_2$ ($b$), in \Thmref{theorem2} and their standard deviations during the pseudo-labeling step of the proposed method.
The green lines represent the sum of the two terms $(a)+(b)$. The results show that asymmetric tri-training minimizes the sum of the two terms (i.e., the upper bound of the ideal loss) during training.
\label{fig:upper_bound}
\vskip 0.12in
\centering
\begin{center}
    \begin{tabular}{c}
        \begin{minipage}{0.48\hsize}
            \begin{center}
                (a) Yahoo! R3
                \includegraphics[clip, width=8.5cm]{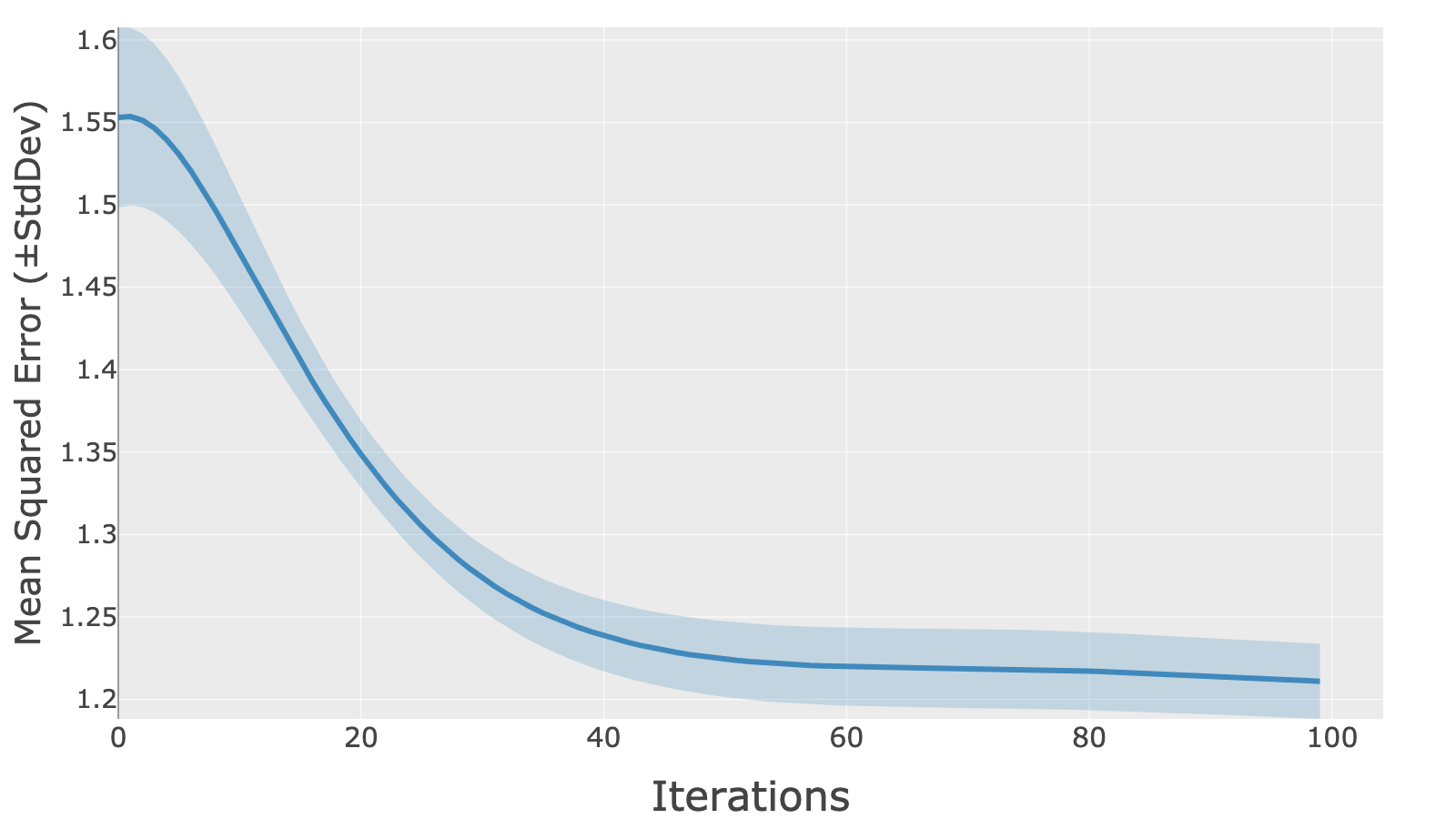}
            \end{center}
        \end{minipage}
        
        \begin{minipage}{0.48\hsize}
            \begin{center}
                (b) Coat
                \includegraphics[clip, width=8.5cm]{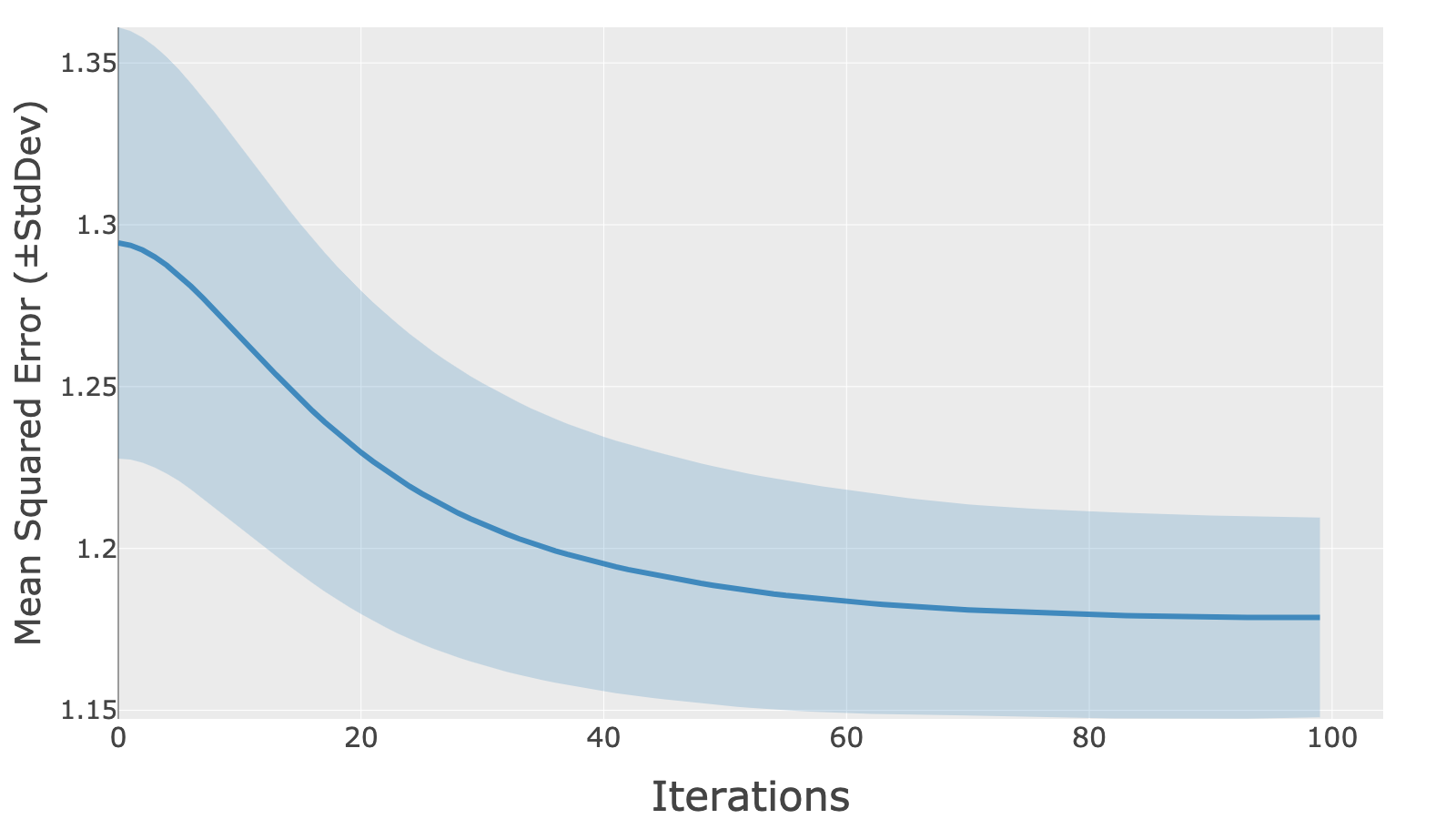}
            \end{center}
        \end{minipage} 
    \end{tabular}
\end{center}
\caption{Improved performance on the test sets by \textit{asymmetric tri-training}}
\vskip 0.05in
\raggedright
\fontsize{9pt}{10pt}\selectfont \textit{Notes}: 
This figure reports averaged MSEs on the test sets and their standard deviations (StdDev) during the pseudo-labeling step of the proposed method. 
The values almost monotonically decrease with iterations. 
These results suggest that minimizing the upper bound of the ideal loss is a valid approach to improve recommenders.
\label{fig:ideal_loss}
\end{figure*}
    
\subsection{Results \& Discussions}
Below, we address the four research questions (RQs). \\

\noindent
\textbf{\textit{RQ1. Is the proposed method robust to the variance problem?}}: 
First, we evaluated the influence of the skewness of the propensity score distribution on the performance of MF-IPS with and without AT using the ML 100K dataset. To evaluate the effects of skewness, we investigated the performance corresponding to varying values of the \textit{min\_items}\footnote{The values of \textit{min\_items} were set to 1, 2, 5, 10, 20, and 50}. A smaller value of \textit{min\_items} introduces a large skewness of the propensity score distribution, as the minimum value of the propensity score in \Eqref{eq:ml_pscore} also becomes small. For example, when \textit{min\_items} is $1$, the minimum relative propensity is $0.0017$, in contrast, when \textit{min\_items} is $50$, the minimum relative propensity is $0.0859$. Note that each model was trained with the specified propensity model in \Eqref{eq:ml_pscore} to evaluate the pure effect of the variance.

\Figref{fig:variance} shows the effect of the skewness of the propensity score distribution on the performance of the MF-IPS with and without AT. The result shows that the MF-IPS without AT is severely affected by the skewness of the propensity distribution, its performance is worsened for a smaller \textit{min\_items}. This is because the IPS approach generally suffers from the variance of the loss function based on the propensity score. In contrast, the MF-IPS with AT performed relatively well, especially when the skewness of the propensity score distribution was large. This is because the final prediction step of our asymmetric tri-training does not rely on the inverse propensity score and thus does not suffer from the variance problem. The result empirically shows that the proposed meta-learning method is robust to the variance problem. \\

\noindent
\textbf{\textit{RQ2. Is the proposed method robust to the choice of propensity score estimator?}}:
Subsequently, we evaluated the influence of the choice of the propensity score estimation model on the performance of the MF-IPS with and without AT. \Tabref{tab:results} summarizes the rating prediction performance evaluated by MSE and MAE and the ranking performance measured by \textit{normalized discounted cumulative gain} (nDCG) \cite{jarvelin2002cumulated} on Yahoo! R3 and Coat datasets.

First, for the Yahoo! R3 dataset, MF-IPS without AT is severely affected by the choice of propensity estimator; only MF-IPS with NB (true) achieves the performance reported in previous works \cite{wang2019doubly,schnabel2016recommendations}, and it completely fails in rating prediction with other propensity estimators. Therefore, MF-IPS is highly susceptible to the propensity misspecification problem. It is difficult to address the effect of selection bias of real-world recommender systems when the MCAR data is unavailable. In contrast, MF-IPS with AT reveals a stable performance with different propensity models and outperforms MF-IPS without AT in most cases. In particular, the proposed method significantly improves the rating prediction accuracies (MSE and MAE) under the realistic situation where the true rating prior is unavailable. Thus, this result validates that the proposed asymmetric tri-training can provide robustness to the choice of the propensity estimation model and improvements of the recommendation quality on biased real-world datasets.

As for the Coat dataset, MF-IPS with and without AT show almost the same rating prediction performance. Moreover, the effect of using different propensity estimators is small. This is because the shift of rating distributions between training and test sets is small in this dataset (see \Figref{fig:rate_dist}). Nonetheless, the proposed method consistently improves the ranking performance (nDCG@3) on this dataset.

In summary, the performance of MF-IPS is substantially affected by the choice of propensity estimators. It is difficult to reveal reasonable performance in most real-world situations when the NB with true prior propensity estimator cannot be used. In addition, the proposed meta-learning method largely improves the recommendation quality especially for the Yahoo! R3 dataset and demonstrates stable performance across different levels of selection bias.  \\

\noindent
\textbf{\textit{RQ3. Does the proposed method actually minimize the upper bound of the ideal loss function?}}: 
Next, we empirically show that the proposed asymmetric tri-training method can actually minimize the upper bound of the ideal loss function derived in \Thmref{theorem2}. 

\Figref{fig:upper_bound} shows the values of the loss on pseudo-labels $(a)$, the similarity between $A_1$ and $A_2$ $(b)$, and their summations $(a) + (b)$ from the pseudo-labeling step of the proposed method. For all datasets, it can be noted that the proposed meta-learning method successfully minimizes the sum of $(a)$ and $(b)$ (the green lines). Thus, as discussed in Section 4.2, the propensity-independent upper bound of the ideal loss function can be minimized effectively using the proposed asymmetric tri-training framework. \\

\noindent
\textbf{\textit{RQ4. Is the upper bound minimization approach valid for minimizing the ideal loss function of the test data?}}: 
Finally, we demonstrate that minimizing the upper bound of the ideal loss function in \Thmref{theorem2} is a valid approach for minimizing the ideal loss function of the test set in \Eqref{eq:ideal}.

\Figref{fig:ideal_loss} shows the MSE on test sets during the pseudo-labeling step of the proposed method. The results suggest that the MSE on the test sets considerably decreases during the pseudo-labeling step; thus, the upper bound minimization approach is empirically justified as an effective way to improve the prediction accuracy from biased explicit feedback.
\section{Conclusion}
In this study, we explored the problem of learning recommenders from MNAR explicit feedback. To this end, we proposed a model-agnostic meta-learning method and demonstrated that it minimizes the part of the propensity-independent upper bound of the ideal loss function, while keeping it tight during training. In the experiments, we empirically demonstrated that the previous propensity-based recommendations are subject to the propensity misspecification and variance issues. Furthermore, we showed that the proposed method is robust to the variance and the choice of the propensity estimation model.

As future work, we plan to apply other unsupervised domain adaptation methods, such as domain adversarial learning \cite{ganin2015unsupervised, ganin2016domain} to the MNAR recommendation. Moreover, we plan to construct a similar learning method for implicit feedback recommendation \cite{liang2016modeling,joachims2017unbiased}. Implicit feedback is prevalent in real-world interactive systems; however, methods for debiasing the implicit feedback recommender have not yet been thoroughly investigated. Thus, we believe that the proposed method can have a significant impact on the implicit feedback recommendation.

\bibliographystyle{ACM-Reference-Format}
\bibliography{atmf.bbl}

\end{document}